\documentclass[a4paper]{jpconf}
\usepackage{graphicx}
\begin{document}

\title{Quantized vortices in two dimensional solid $^4$He}

\author{M Rossi, D E Galli, P Salvestrini and L Reatto}

\address{Dipartimento di Fisica, Universit\`a degli Studi di Milano, 
        via Celoria 16, 20133 Milano, Italy}

\ead{luciano.reatto@unimi.it}

\begin{abstract}
Diagonal and off-diagonal properties of 2D solid $^4$He systems doped with a
quantized vortex have been investigated via the Shadow Path Integral Ground State
method using the fixed-phase approach.
The chosen approximate phase induces the standard Onsager-Feynman flow field.
In this approximation the
vortex acts as a static external potential and the resulting Hamiltonian can be
treated exactly with Quantum Monte Carlo methods.
The vortex core is found to sit in an interstitial site and a very weak relaxation of
the lattice positions away from the vortex core position has been observed.
Also other properties like Bragg peaks in the static structure factor or the behavior
of vacancies are very little affected by the presence of the vortex.
We have computed also the one-body density matrix in perfect and defected $^4$He
crystals finding that the vortex has no sensible effect on the off-diagonal long
range tail of the density matrix.
Within the assumed Onsager Feynman phase,
we find that a quantized vortex cannot auto-sustain itself
unless a condensate is already present like when dislocations are present.
It remains to be investigated if backflow can change this conclusion.
\end{abstract}

Quantized vortices are one of the most genuine manifestation of the presence of 
superfluidity in many body quantum systems, but from the microscopic point of view
no complete understanding of them has been reached yet.
Recently quantized vortices have been related to the supersolidity issue \cite{bali};
in fact, arguments in favor of a vortex phase in low
temperature solid $^4$He, preceding the supersolid transition, are appeared in
literature \cite{and1,and2,kubo}.
With respect to this possible connection one of the fundamental questions to be
answered is: what does a quantum vortex look like in solid Helium from a microscopic
point of view?

Dealing with vortices is a really hard task for microscopic methods, and it
calls for some approximations or assumptions 
[5-10].
In fact, the wave function has to be an eigenstate of the angular momentum, so
it needs a phase. 
Following the well established routine for the ground state, once chosen a 
variational ansatz for the wave function, one could be tempted to correct it by 
means of exact zero temperature Quantum Monte Carlo (QMC) techniques.
Unfortunately this is actually not possible because of the sign problem that affects 
QMC methods.
The most followed recipe is to improve the variational description via QMC, but 
releasing the exactness of the methods in favor of approximations that allow to 
avoid the sign problem, like for example fixed phase \cite{orti} or fixed nodes 
\cite{gior}.

We study here the properties of a single vortex in solid $^4$He via the Shadow-PIGS
(SPIGS) method with fixed phase approximation.
The many-body wave function can be written as $\Psi(R) = e^{i\Omega(R)}\Psi_0(R)$,
where $\Omega(R)$ is a many-body phase, $\Psi_0(R)$ is the modulus of the wave 
function and $R=(\vec r_1, \vec r_2, \dots, \vec r_N)$ are the coordinates of the 
$N$ particles.
$\Psi(R)$ describes a quantum state of the system if it is a solution of the 
time independent Schr\"odinger equation: from $\hat H\Psi(R) = E\Psi(R)$ it is possible to obtain two 
coupled differential equations for $\Omega(R)$ and for $\Psi_0(R)$.
The fixed phase approximation consists in assuming the functional form of $\Omega (R)$
as given and to solve the equation
\begin{equation}
\label{schreff}
 \left[-\frac{\hbar^2}{2m}\sum_{i=1}^N\nabla^2_i + V(R) + 
        \frac{\hbar^2}{2m}\sum_{i=1}^N(\vec\nabla_i\Omega)^2\right]\Psi_0(R) = E\Psi_0(R).
\end{equation}
for $\Psi_0(R)$.
Solving (\ref{schreff}) is equivalent to solve the original time independent
Schr\"odinger equation for the 
$N$-particle with an extra potential term 
$V_{\rm e}=\frac{\hbar^2}{2m}\sum_{i=1}^N(\vec\nabla_i\Omega)^2$.

The simplest choice for the phase is the well known Onsager-Feynman (OF)
phase \cite{feyn}:
$\Omega(R) = l\sum_{i=1}^N\theta_i$ 
(where $\theta_i$ is the angular polar coordinate of the $i$-th particle).
$\Psi(R)$ is an eigenstate of the $z$ component of the angular momentum operator $\hat L_z$ with 
eigenvalue $\hbar Nl$, being $l=1,2,\dots$ the quantum of circulation.
This choice for $\Omega(R)$ gives rise to the standard OF flow field: in fact
the extra-potential in (\ref{schreff}) reads 
\begin{equation}
\label{pot}
V_{\rm e}=\frac{l^2\hbar^2}{2m}\sum_{i=1}^N\frac{1}{r_i^2}
\end{equation}
where $r_i$ is the radial polar coordinate of the $i$-th particle.

In order to sustain a quantized vortex, the system should display a macroscopic phase
coherence, and at $T=0$ K this means that solid $^4$He should house a 
Bose-Einstein condensate (BEC).
It is known from QMC results that no BEC is present in the perfect crystal 
[12-15], but if the vortex turns out to be able to induce a BEC 
it could be a self-sustaining excitation.
On the other hand, it is largely accepted that defects are able to induce BEC \cite{disl}, 
and then a defected crystal can safely sustain a quantized vortex.
Here we report on the study of a two dimensional (2D) $^4$He crystal with and without dislocations.
In fact, dislocations can be included in the 2D crystal without imposing boundary
constraints \cite{disl}.
Moreover the 2D system allows to reach large distances keeping the number of 
particle in the simulation at a tractable level, and this is a desirable feature
when interested in off-diagonal properties of the system.

We face the task of solving (\ref{schreff}) with the extra-potential given by (\ref{pot})
when $l=1$ with the SPIGS method \cite{spigs1,spigs2}, which allows to obtain the lowest 
eigenstate of a given Hamiltonian by projecting in imaginary time a SWF \cite{vit1} taken 
as trial wave function.
The SPIGS method is unbiased by the choice of the trial wave function and the only
inputs are the interparticle potential and the approximation for the imaginary time
propagator \cite{pata}.
As He-He interatomic potential we have considered the HFDHE2 Aziz potential \cite{aziz} 
and we have employed the pair-Suzuki approximation \cite{pata} for the imaginary time
propagator with time step $\delta\tau = 1/120$ K$^{-1}$.
One difficulty with (\ref{pot}) is that the potential is long range so that either one
puts the system in a bucket \cite{orti,vit2} or one should consider a vortex lattice \cite{sadd}.
Such complications can be avoided by multiplying $1/r^2$ in (\ref{pot})
by a smoothing function 
\begin{equation}
\label{smooth}
 \chi(r) = \left\{
 \begin{array}{ll}
  1 & r<\Delta\\
  \e^{-(r-\Delta)^2/(r-L/2)^2} & \Delta\le r\le L/2 \\
  0 & r> L/2
 \end{array}
 \right.
\end{equation}
($L$ being the side of the simulation box) so that standard periodic boundary conditions
can be applied.
With this choice, the extra-potential is equivalent to the OF one only for $r<\Delta$,
the provided $\Psi_0$ is no more an exact eigenstate of $\hat L_z$ but it is close to it 
in the interesting region of the vortex core if $\Delta$ is large enough.
Here we have used $\Delta=30$\AA.
We have performed simulations at $\rho=0.0765$\AA$^{-2}$ in a nearly squared box designed
to house a perfect triangular crystal with $M=572$ lattice sites, and a crystal with 
10 vacancies ($N=562$); such vacancies in the initial configuration transform themselves in
dislocations \cite{disl}.

\begin{figure}
\includegraphics*[width=7cm]{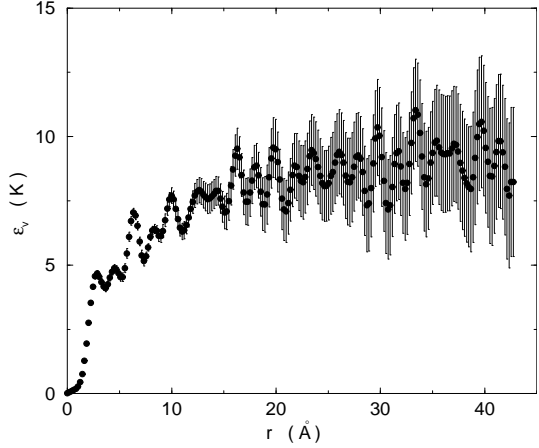}\hspace{2pc}%
\begin{minipage}[b]{7cm}\caption{\label{fig:e2D} Integrated vortex energy $\varepsilon_v$ 
                                 with error bars
                                 as a function of the distance from the core in perfect 2D 
                                 solid $^4$He at $\rho=0.0765$\AA$^{-2}$.}
\end{minipage}
\end{figure}
In Fig.~\ref{fig:e2D} we report our results for the integrated vortex energies
$\varepsilon_v(r) = (E_v(r)-E(r))/N$ ($E_v(r)$ and $E(r)$ are, respectively, the energy of the 
particles that lie inside the disk of radius $r$ in the system with and without vortex) as a 
function of the distance from the core in the perfect crystal.
The center of mass of the system is not fixed,
and we find that, independently on the starting 
configuration of the crystal, the vortex core sits in an interstitial site.
We also find a very small relaxation of the surrounding lattice around the vortex core.

\begin{figure}
\includegraphics*[width=7cm]{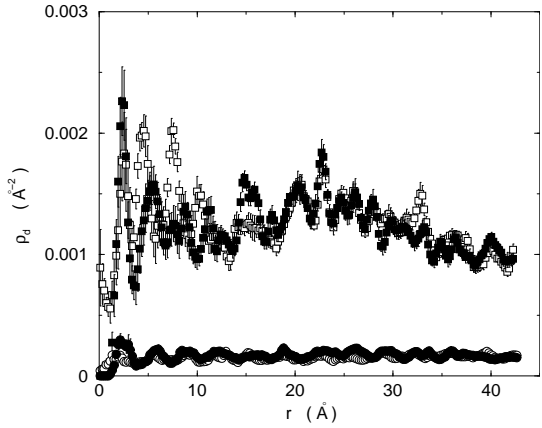}\hspace{2pc}%
\begin{minipage}[b]{7cm}\caption{\label{fig:x2D} Comparison between the radial defect 
                                 distribution $\rho_d(r)$ as a function of the distance from
                                 the origin (vortex core when vortex is present) for a 2D
                                 solid $^4$He at $\rho=0.0765$\AA$^{-2}$ with (filled symbols) 
                                 and without (open symbols) a vortex in the perfect (circles)
                                 and in defected (squares) crystal.}
\end{minipage}
\end{figure}
In order to study the effects of the vortex on the crystal properties we have monitored 
the static structure factor, the pair distribution function and the radial defect distribution
$\rho_d(r)$ (i.e. the distribution of the particle whose coordination is different from 6 as a 
function of the distance from the origin where the vortex core is located).
In Fig.~\ref{fig:x2D} we plot our results for $\rho_d(r)$ both for the perfect and for the
defected crystal.
We find that in the defected case the radial defect distribution is about an order of 
magnitude larger than in the perfect one; however, the results of the system with and
without vortex are very close each other so that we conclude that the OF vortex does not affect 
in a sensible way the disorder which behaves as in the system without vortex.
We come at a similar conclusion for the crystalline structure, since both the static structure 
factor and the pair distribution function show no appreciable differences for the system
with and without the vortex.

\begin{figure}
\includegraphics*[width=7cm]{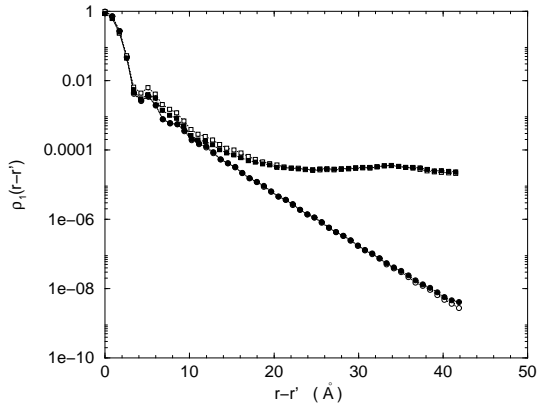}\hspace{2pc}%
\begin{minipage}[b]{7cm}\caption{\label{fig:c2D} One body density matrix $\rho_1$
                                 in 2D solid $^4$He without (open symbols) and
                                 with (filled symbols) a vortex for the perfect crystal
                                 (circles) and for the crystal with dislocations (squares).}
\end{minipage}
\end{figure}
We have computed also the one-body density matrix $\rho_1$ for the perfect and for the 
defected crystal with and without the vortex in order to investigate the vortex effects on
the off-diagonal properties.
Our results are reported in Fig.~\ref{fig:c2D}.
$\rho_1$ for the crystal with vortex are indistinguishable within
the error bars from the ones obtained without vortex.
We conclude that the OF vortex is not able to induce a Bose-Einstein condensation (BEC) 
in the perfect crystal, or to increase the already present condensate fraction in the 
defected one.

Since no BEC is present in the perfect crystal and the OF vortex is not able to induce
it by itself, we can conclude that perfect 2D solid $^4$He can not sustain vortices of the OF type.
Thus the OF wave function is a possible representation of a vortex in solid 
$^4$Helium only when BEC is already present, like in a defected crystal.
Preliminary results in three dimensional solid $^4$He seem to confirm such conclusions
for the OF vortex.

In the liquid phase, the OF phase \cite{feyn} has been improved with the 
inclusion of back-flow (BF) correlations \cite{orti,sadd}.
The effect of backflow increases \cite{sadd} at higher densities and it might well
become dominant in the solid phase.
Computations along this line are in progress.
Since the BF terms acts mainly near the vortex core, we might expect that BF will not modify
much the off-diagonal properties in 2D, where the vortex core is a point defect,
while it could become relevant in 3D where the vortex core is an extended defect.

This work was supported by Regione Lombardia and CILEA Consortium through a LISA 
Initiative (Laboratory for Interdisciplinary Advanced Simulation) 2010 grant 
[link:http://lisa.cilea.it].

\end{document}